\documentclass[amsmath, amssymb, reprint,showkeys,aps,pra,twocolumn]{revtex4-2}

\usepackage{color} 
\usepackage{epsfig}
\usepackage{soul}
\usepackage{hyperref}
\usepackage{slashed}
\usepackage{graphicx}
\usepackage{amsmath}
\usepackage{latexsym}
\usepackage{epstopdf}
\usepackage{amsmath}
\usepackage{enumitem}
\usepackage{amssymb,amsmath}
\usepackage{multirow}
\usepackage{mathtools}
\usepackage{bbm}
\usepackage{bm}
\usepackage{amsfonts}
\usepackage{amssymb}
\usepackage{calligra}
\usepackage{calrsfs}
\usepackage{makecell}
\usepackage[normalem]{ulem}
\usepackage{braket}
\usepackage[british,UKenglish,USenglish,american]{babel}
\usepackage{microtype}

\begin{document}

\title{Scattering theory and equation of state of a spherical two-dimensional Bose gas}

\author{A. Tononi}
\email{andrea.tononi@universite-paris-saclay.fr}
\affiliation{Universit\'e Paris-Saclay, CNRS, LPTMS, 91405 Orsay, France}
\affiliation{Dipartimento di Fisica e Astronomia ``Galileo Galilei", Universit\`a di Padova, via Marzolo 8, 35131 Padova, Italy}
\affiliation{Istituto Nazionale di Fisica Nucleare (INFN), Sezione di Padova, 
via Marzolo 8, 35131 Padova, Italy}


\begin{abstract}
We analyze the scattering problem of identical bosonic particles confined on a spherical surface. 
At low scattering energies and for a radius much larger than the healing length, we express the contact interaction strength in terms of the $s$-wave scattering length. 
Adopting this relation, we are then able to regularize the zero-point energy of the spherical Bose gas and to obtain its equation of state, which includes the corrections due to the finite radius of the sphere and coincides with the flat-case result in the infinite-radius limit. 
We also provide a microscopic derivation of the superfluid density of the system, reproducing a result postulated in a previous work. 
Our results are relevant for modeling the ongoing microgravity experiments with two-dimensional bubble-trapped Bose-Einstein condensates. 
\end{abstract}

\maketitle

\section{Introduction} 
The main theoretical advances in the study of weakly-interacting Bose gases were achieved by reducing the complexity of the full many-body problem to the analysis of the quantum statistical properties of few noninteracting constituents.  
In the ideal Bose gas, for instance, one can derive the system thermodynamics by solving the Schr\"odinger equation of a single boson  \cite{bose1924,einstein1924}, while a zero-temperature dilute gas is well described by a single macroscopic field satisfying the mean-field Gross-Pitaevskii equation \cite{gross1961,pitaevskii1961}. 
The explicit many-body nature of a quantum gas emerges when considering the fluctuations beyond the mean-field configuration, but, nonetheless, these are rephrased in terms of noninteracting Bogoliubov quasiparticles \cite{landau1941,bogoliubov1947}. 
Coherently with these examples, the interatomic interactions of a bosonic gas, which are the object of our investigation, are also typically modeled through the analysis of a single particle scattering on the potential core. 

Typically, since the precise form of the interaction potential is unknown, a beyond-mean-field description of a Bose gas is based on the assumption that the bosons interact with a zero-range two-body interaction of strength $g_0$. 
This physically motivated, but nonetheless rough, assumption yields a divergence of the zero-point energy of the quasiparticles and requires proper regularization techniques, such as dimensional regularization \cite{andersen2004}, the inclusion of momentum cutoffs \cite{salasnich2016}, and the use of convergence factors \cite{altland2010}. 
After the regularization, the analysis of the low-energy scattering properties of two bosonic particles allows us to relate the interaction strength $g_0$ with the $s$-wave scattering length $a_s$, which is a measurable quantity. 
This can be done within the Born approximation in three-dimensional weakly interacting systems, and
these steps, therefore, yield the equation of state expressed in terms of physical parameters, which can be compared with the experiments. 

In the two-dimensional (2D) case, which will be analyzed in the present paper, the general procedure described above requires the summation of the Born series, and has been implemented several times to obtain the equation of state (see Refs.~\cite{schick1971,popov1972,cherny2001,mora2003,pricoupenko2004,petrov2016}).
These works, however, deal with a uniform, infinite, and flat gas. 
In this work, instead, we consider a finite-size curved Bose gas confined on the surface of a sphere \cite{tononi2019}, and we derive its regularized equation of state by implementing the scattering theory in this context \cite{zhang2018}. 

The present study supplements the work of Ref.~\cite{tononi2021} and is motivated by the ongoing microgravity experiments with shell-shaped Bose-Einstein condensates \cite{carollo2021,lundblad2019}. 
This geometry is engineered via a magnetic trapping of the atoms with radiofrequency-induced adiabatic potentials \cite{zobay2001}, and requires microgravity conditions due to the technical difficulty of counterbalancing gravity in ground-based experiments \cite{colombe2004,white2006,arazo2021}. 
By tuning the shell thickness, these experiments will allow the tuning of the effective two-body interaction, exploring different interaction regimes that are quantitatively described by our equation of state.
Moreover, future analyses will enable the study of vortex physics \cite{padavic2020,bereta2021,turner2010}, of the hydrodynamic excitations \cite{lannert2007,sun2018,padavic2018}, of the shell thermodynamics \cite{rhyno2021}, and of the Berezinskii-Kosterlitz-Thouless transition \cite{tononi2021,he2021}. 

A brief synopsis of the paper is as follows. 
After modeling the zero-range scattering of two particles on a spherical surface (Sec.~\ref{IIA}), we link the low-energy scattering amplitude with the $s$-wave scattering length of the real interatomic potential (Sec.~\ref{IIB}). 
With this procedure, we are able to regularize the zero-point energy of the spherical Bose gas (Sec.~\ref{IIIA}), obtaining the equation of state at zero temperature and at finite temperature. 
Our formalism also allows us to provide a microscopic derivation of the superfluid density of the system (Sec.~\ref{IIIB}), which we calculate from the nonclassical moment of inertia of a rotating sphere.

\section{Scattering on a spherical surface} 
\label{II} 
The scattering problem of identical particles of mass $m$ can be considerably simplified when it is possible to analyze independently the center-of-mass motion and the relative motion.
When this reduction is implemented, the relative problem is usually formulated as a Schr\"odinger equation for a single particle with reduced mass $m/2$ that scatters in the two-body interaction potential $\hat{V}$. 
If we consider two interacting particles on the surface of a sphere, however, it is not generally valid to assume that the center of mass motion and the relative one are decoupled.
In this case, a possible work-around consists of discussing the scattering of a single particle with mass $m$ in the interaction potential \cite{zhang2018}. 
Another possibility of simplification that we will implement in this paper consists of assuming that the radius of the sphere $R$ is much larger than the healing length $\xi=\sqrt{\hbar^2/(2m\mu)}$, with $\mu$ being the chemical potential of the many-particle system, and assuming that the relative wave vector of the interacting particles is much larger than the inverse radius. 
Under these conditions, scattering occurs in a near-Euclidean regime; that is, there are perturbative corrections with respect to the scattering problem in the 2D flat case \cite{zhang2018}, and the problem can be approximately formulated in terms of a particle with reduced mass moving in the interaction potential.
It is relevant to note that, even for particles interacting in a finite-size 2D box, the separability occurs only when the box size exceeds both $\xi$ and the inverse of the relative wave vector, similar to our assumption in the spherical case. 

By following the analysis of Lippmann and Schwinger \cite{lippmann1950}, it is possible to reformulate the Schr\"odinger equation describing the dynamics of the particles in a time-independent formalism, which requires the calculation of probability amplitudes between the initial states of the system, i.~e., $\ket{\phi}$, and the outgoing scattering states, i.e., $\ket{\Psi^{(+)}}$. 
The former are usually assumed to be the eigenstates of the noninteracting Hamiltonian $\hat{H}_0$, with eigenvalues $\mathcal{E}_{0}$, while the scattering state solves the Schr\"odinger equation for the complete Hamiltonian $\hat{H}_0+\hat{V}$.
Introducing the $\hat{\mathcal{T}}$-matrix, which satisfies the relation $\hat{\mathcal{T}} \ket{\phi} = \hat{V} \ket{\Psi^{(+)}}$, one finds the Lippmann-Schwinger equation \cite{lippmann1950,stoof}
\begin{equation}
\hat{\mathcal{T}} = \hat{V} + \hat{V} \, \frac{1}{\mathcal{E}_{0}-\hat{H}_0 +i \eta} \, \hat{\mathcal{T}}, 
\label{Tsphere}
\end{equation}
which admits an iterative solution for $\hat{\mathcal{T}}$ that generates the Born series \cite{stoof}, and where $\eta \to 0^+$ is a small positive parameter that is added to regularize the denominator. 

In the following, we calculate the $\hat{\mathcal{T}}$-matrix elements for low-energy scattering on the surface of a sphere. 
We will then link this result with a measurable scattering parameter, the $s$-wave scattering length, which will allow us to obtain the regularized equation of state of a Bose gas on the surface of a sphere. 

\subsection{Low-energy solution of the Lippmann-Schwinger equation}
\label{IIA}
To obtain the explicit solution of the Lippmann-Schwinger equation, we consider a spherical surface with radius $R$, parametrized by a system of spherical coordinates $\{\theta,\varphi\} \in [0,\pi]\times[0,2\pi]$, so that $\ket{\theta,\varphi}$ is the position vector. 
In our analysis, the radius determines a natural energy scale $E_R= \hbar^2/(mR^2)$, with $\hbar$ being the Planck's constant, which we use to rescale all the energies and Hamiltonians of the present section. 

The scattering state $\ket{\phi}=\ket{l_0,m_{l_0}}$ of a particle with reduced mass satisfies the following adimensional Schr\"odinger equation:  
\begin{equation}
\hat{H}_0 \ket{l_0,m_{l_0}} = \mathcal{E}_{l_0} \ket{l_0,m_{l_0}},
\end{equation}
where the dimensionless Hamiltonian $\hat{H}_0$ (expressed in units of $E_R$) reads
\begin{equation}
\hat{H}_0 = - \bigg[  \frac{1}{\sin \theta} \frac{\partial}{\partial \theta} \bigg( \sin \theta \frac{\partial}{\partial \theta}  \bigg) + \frac{1}{\sin^2 \theta} \frac{\partial^2}{\partial \varphi^2}\bigg], 
\label{schrodingereq}
\end{equation}
and is thus proportional to the angular momentum operator in spherical coordinates $\hat{L}^2$. 
The dimensionless energies $\mathcal{E}_{l_0}$ are given by 
\begin{equation}
\mathcal{E}_{l_0} = l_0(l_0+1),
\end{equation}
and the noninteracting eigenstates $\ket{l,m_l}$ are such that $\mathcal{Y}_l^{m_l}(\theta,\varphi) = \braket{\theta,\varphi | l,m_l}$ are the spherical harmonics, with $l=0,1,2,...$, and $m_l = -l,...,l$ being the quantum numbers of the angular momentum.  
We emphasize that the bases $\{\ket{\theta,\varphi}\}$ and $\{\ket{l,m_l}\}$ are both orthonormal since they satisfy the identities 
\begin{align}
\braket{\theta',\varphi'|\theta,\varphi} &= \delta(\cos\theta -\cos \theta')\, \delta(\varphi-\varphi'),
\\
\braket{l',m_l'|l,m_l} &= \delta_{l,l'} \, \delta_{m_l,m_l'},
\end{align}
where $\delta$ denotes both the Dirac and Kronecker deltas, and they are also complete since 
\begin{align}
1 &= \int_0^{2\pi} \text{d}\varphi \int_{0}^{\pi}  \text{d}\theta \sin \theta \, \ket{\theta,\varphi} \bra{\theta,\varphi},
\label{identity1}
\\
1 &= \sum_{l=0}^{\infty} \sum_{m_l=-l}^{l} \, \ket{l,m_l} \bra{l,m_l}.
\label{identity2}
\end{align}

To solve the Lippmann-Schwinger equation \eqref{Tsphere} and thus to solve the scattering problem on the sphere, we need to model explicitly the two-body potential $\hat{V}$. 
A widely used approximation in the field of weakly interacting quantum gases consists of describing the low-energy isotropic interactions via a zero-range two-body potential $\hat{V}_0$. 

In a flat system, in which the wave vectors are a good basis, this assumption corresponds to setting $\hat{V}_0 = \tilde{g}_0 \, \delta(\mathbf{r})$, so that the matrix element between different wave vector states is constant and equal to the contact interaction strength $\tilde{g}_0$, namely $\braket{\mathbf{k}' |\hat{V}_0 | \mathbf{k}_0} = \tilde{g}_0$. 
We model the interaction in the spherical case under the same assumption, by setting 
\begin{equation}
\hat{V}_0 =  V_0(\theta,\varphi),
\label{zerorangeint}
\end{equation}
where $V_0(\theta,\varphi) = \tilde{g}_0 \, \delta(1 -\cos \theta)\, \delta(\varphi)$, and $\tilde{g}_0 = g_0 m /\hbar^2$ represents the zero-range interaction strength between the particles, which have a relative angular distance corresponding to the coordinates $(\theta,\varphi)$.
In contrast to flat systems, it must then be stressed that the matrix element between different spherical harmonics states is not constant.
We find, in particular, that
\begin{equation}
\braket{l',m_{l'}=0 | \hat{V}_0 | l_0,m_{l_0}=0} = \tilde{g}_0 \, \frac{\sqrt{(2l'+1)(2l_0+1)}}{4\pi},
\label{potential}
\end{equation}
which can be verified by inserting the identity of Eq.~\eqref{identity1} inside the bracket at the first side and using the definition of spherical harmonics.
In this expression, as we will assume in the following, we have set $m_{l'}=0=m_{l_0}$, which corresponds to describing only $s$-wave scattering, in which the interacting particles do not exchange quanta of the angular momentum. 
This assumption is valid for low scattering energies, at which $\braket{l',m_{l'} | \hat{V} | l_0,m_{l_0}} \approx \braket{l',m_{l'}=0 | \hat{V}_0 | l_0,m_{l_0}=0}$.
Before proceeding, we note an important fact implicit in Eq.~\eqref{potential}: in the context of the approximations done, the ratio
\begin{equation}
\tilde{g}_0 = \frac{4\pi\braket{l',m_{l'}=0 | \hat{V}_0 | l_0,m_{l_0}=0}}{\sqrt{(2l'+1)(2l_0+1)}}
\label{g0bare}
\end{equation}
coincides with the bare interaction strength $\tilde{g}_0$.

Let us substitute the potential $\hat{V}$ with $\hat{V}_0$ in the Lippmann-Schwinger equation \eqref{Tsphere}, which becomes 
\begin{equation}
\hat{\mathcal{T}} = \hat{V}_0 + \hat{V}_0 \, \frac{1}{\mathcal{E}_{l_0}-\hat{H}_0 +i \eta} \, \hat{\mathcal{T}}, 
\label{tsphere0}
\end{equation}
and then we calculate the $\mathcal{T}$-matrix element $\mathcal{T}_{l',l_{0}}=\bra{l',m_l'=0}\hat{\mathcal{T}}\ket{l_0,m_{l_0}=0}$, obtaining
\begin{align}
\mathcal{T}_{l',l_{0}} =& \tilde{g}_0 \, \frac{\sqrt{(2l'+1)(2l_0+1)}}{4\pi} \times
\label{result} \\
 &\bigg[ 1 +\sum_{l=0}^{\infty} \sum_{m_l=-l}^{l} \frac{\sqrt{2l+1}}{\sqrt{2l_0+1}}
\frac{\bra{l,m_l}\hat{\mathcal{T}}\ket{l_0,m_{l_0}=0}}{\mathcal{E}_{l_0}-\mathcal{E}_{l}+i\eta}\bigg],
\nonumber
\end{align}
where we have inserted the identity of Eq.~\eqref{identity2} at the right-hand side. 
Note that this integral equation for $\mathcal{T}_{l',l_{0}}$ is not closed, since its solution requires the knowledge of the matrix element $\bra{l,m_l}\hat{\mathcal{T}}\ket{l_0,m_{l_0}=0}$ for $m_l \neq 0$. 
However, consistently with our assumption \eqref{potential} for the interaction potential, we neglect all the $\mathcal{T}$-matrix elements with $m_l \neq 0$, and we solve the equation 
\begin{align}
\mathcal{T}_{l',l_{0}} =& \tilde{g}_0 \, \frac{\sqrt{(2l'+1)(2l_0+1)}}{4\pi} \times 
\label{bornTll0}\\
 &\bigg[ 1 +\sum_{l=0}^{\infty} \frac{\sqrt{2l+1}}{\sqrt{2l_0+1}}
\frac{\mathcal{T}_{l,l_{0}}}{\mathcal{E}_{l_0}-\mathcal{E}_{l}+i\eta}\bigg],
\nonumber
\end{align}
which is a closed ``integral'' equation for $\mathcal{T}_{l',l_{0}}$. 

Following the usual operatorial approach of scattering theory, Eq.~\eqref{bornTll0} can be solved by iteration (i.e. substituting repeatedly $\mathcal{T}_{l,l_{0}}$ appearing at the right-hand side with the whole right-hand side) and summing the resulting geometric series \cite{stoof}. 
This procedure leads to 
\begin{equation}
\frac{\sqrt{(2l'+1)(2l_0+1)}}{4\pi\mathcal{T}_{l',l_{0}}} = \frac{1}{\tilde{g}_0} + \frac{1}{4\pi} \sum_{l=0}^{\infty} \frac{(2l+1)}{\mathcal{E}_l-\mathcal{E}_{l_0}-i\eta},
\end{equation}
in which the right-hand side does not depend on $l'$, $m_{l'}$. 
Due to the choice of a zero-range interaction, indeed, the calculation is actually valid in a regime of very low-energy scattering in which the interaction strength is $\tilde{g}_0 \ll 1$, and the only relevant $\mathcal{T}$-matrix elements are those for which $\bra{l',m_l'=0}$ is near to $\ket{l_0,m_{l_0}=0}$, and the matrix element $\mathcal{T}_{l',l_{0}}$ coincides with the on-shell $\mathcal{T}$-matrix $\mathcal{T}(\mathcal{E}_{l_0}+i\eta)$ \cite{pricoupenko2011}. 
In analogy with Eq.~\eqref{g0bare}, we introduce the quantity 
\begin{equation}
\tilde{g}_e (\mathcal{E}_{l_0}+i\eta) = \frac{4\pi \mathcal{T}_{l',l_{0}}}{\sqrt{(2l'+1)(2l_0+1)}}.
\end{equation}
In this context, we can interpret the effective coupling constant $\tilde{g}_e (\mathcal{E}_{l_0}+i\eta)$ as a ``renormalized'' interaction strength that includes the multiple zero-range scattering processes generated by the iterative solution of Eq.~\eqref{bornTll0}.
Thanks to the definition of $\tilde{g}_e$, we find a familiar equation in the context of scattering theory \cite{salasnich2016}, namely
\begin{equation}
\frac{1}{\tilde{g}_e (\mathcal{E}_{l_0}+i\eta)} = \frac{1}{\tilde{g}_0} + \frac{1}{4\pi}\sum_{l=0}^{\infty} \frac{2l+1}{\mathcal{E}_l-\mathcal{E}_{l_0}-i\eta},
\label{TEnocutoff}
\end{equation}
which relates the effective interaction strength $\tilde{g}_e$ to the bare interaction strength $\tilde{g}_0$.
This equation, which goes beyond the Born approximation of setting $\mathcal{T}_{l,l_{0}} \approx \braket{l,m_{l}=0 | \hat{V}_0 | l_0,m_{l_0}=0}$ in Eq.~\eqref{bornTll0}, is a standard one for zero-range scattering \cite{dalibard}, and is a crucial step for obtaining the equation of state of a 2D quantum gas \cite{mora2003}.

To obtain an explicit analytical result for $\tilde{g}_e$, we need to calculate the sum on the right-hand side of Eq.~\eqref{TEnocutoff}. We rewrite this term as 
\begin{equation}
\int_{0}^{l_c} \text{d}l \, \frac{(2l+1)}{\mathcal{E}_l-\mathcal{E}_{l_0}-i\eta},
\label{int}
\end{equation}
where we included a finite cutoff $l_c$ to avoid obtaining a divergent result. 
The cutoff $l_c$ is finite but is much larger than $l_0$, and it is regulated by physical processes that are not described by the simple zero-range interaction $\hat{V}_0$ that we have assumed. 
The next equations should be thought of in the limit of $l_c \gg 1$, but we stress that, in the final expression for the equation of state, this cutoff will cancel out and the limiting procedure will be trivial. 
We calculate Eq.~\eqref{int} expressing the integrand as $1/[h(l)-i\eta] = [h(l)+i\eta]/[h(l)^2+\eta^2]$, and we finally obtain $\tilde{g}_e (\mathcal{E}_{l_0})$ as 
\begin{equation}
\tilde{g}_{e,l_c}(\mathcal{E}_{l_0}) = - \frac{2\pi}{-\frac{2\pi}{\tilde{g}_0} + \frac{1}{2}\ln\big[ \frac{l_0(l_0+1)}{l_c(l_c+1)} \big] -i\frac{\pi}{2} },
\label{Tlc}
\end{equation}
where we took the limit $\eta \to 0^+$ at the end and where the subindex $l_c$ reminds us of the inclusion of the ultraviolet cutoff $l_c$. 
Now that $\tilde{g}_{e,l_c} (\mathcal{E}_{l_0})$ is known, let us state again our main goal: we want to express the interaction strength $\tilde{g}_0$ as a function of the cutoff $l_c$, of $l_0$, and of the $s$-wave scattering length. 
To obtain this relation, we will impose that
\begin{equation}
\tilde{g}_{e,l_c} (\mathcal{E}_{l_0}) = f_0(\mathcal{E}_{l_0}), 
\label{condition}
\end{equation}
where $f_0(\mathcal{E}_{l_0})$ is the scattering amplitude of the partial $s$ wave for an incident wave with energy $\mathcal{E}_{l_0}$, and $\tilde{g}_{e,l_c} (\mathcal{E}_{l_0})$ is essentially a renormalized interaction strength \cite{note2}. 
In the next section, {inspired by the analysis of potential scattering on a spherical surface of Ref.~\cite{zhang2018}, we calculate $f_0(\mathcal{E}_{l_0})$.}

\subsection{The \textit{s}-wave scattering amplitude}
\label{IIB}
The following calculations are mainly based on the analysis of Ref. \cite{zhang2018}, which discussed the potential scattering of a single particle of mass $m$ on the surface of a sphere: here we implement an analogous calculation for a particle with reduced mass $m/2$. 
Note that the use of the reduced mass is implicitly included with our rescaling in terms of the energy $E_{R}\equiv \hbar^2/[2(m/2)R^2]$. 
As discussed in Section \ref{II}, this is a valid approximation when the radius of the sphere is much larger than the healing length of the system, and if the relative scattering energy is much larger than $E_{R}$ (which, since we rescale the energies with $E_{R}$ itself, means $\mathcal{E} \gg 1$). 

To define the $s$-wave scattering length, we go back to the scattering problem for the real interatomic potential $\hat{V}$. 
Let us assume that $\hat{V}$ is a spherically-symmetric potential, in which the range of the two-body interaction $r_0 = R \, \theta_0$, with $\theta_0$ being the angular range, is much smaller than the radius of the sphere $R$. 
In this case, at larger angular distances with respect to $\theta_0$, the scattering problem can be expressed as \cite{zhang2018}
\begin{equation}
\hat{H}_0 \Psi_{\nu}^{\mu}(\theta,\varphi) = \mathcal{E}_{\nu} \Psi_{\nu}^{\mu}(\theta,\varphi) ,\qquad \theta > \theta_0,
\label{scatteringproblemas}
\end{equation}
which, formally, coincides with the free Schr\"odinger equation. 
However, the eigenfunctions $\Psi_{\nu}^{\mu}$ are in general different from the spherical harmonics $\mathcal{Y}_l^{m_l}$, {since $\nu$ is a continuous index,} and in the region inside the interaction range, i.~e., $\theta \leq \theta_0$, these eigenfunctions must be matched with the solution of the interacting eigenproblem. 

The general solution of Eq.~\eqref{scatteringproblemas} for $s$-wave scattering {(for which $\mu=0$),} reads \cite{NIST}
\begin{equation}
{
\Psi_{\nu}^{0}(\theta,\varphi) = A  P_{\nu}^{0}(\cos \theta) + B Q_{\nu}^{0}(\cos \theta) ,\quad \theta> \theta_0
}
\end{equation}
where $P_{\nu}^{0}(\cos \theta)$ and $Q_{\nu}^{0}(\cos \theta)$ are the associated Legendre functions of first and second kinds for $\mu=0$ and $A$ and $B$ are coefficients. 
By properly redefining these quantities, the solution can be put also in the form 
\begin{equation}
\Psi_{\nu}^{0}(\theta,\varphi) \propto  P_{\nu}^{0}(\cos \theta) + \frac{f_0(\mathcal{E}_{\nu})}{4 i} \bigg[  P_{\nu}^{0}(\cos \theta) + \frac{2i}{\pi} Q_{\nu}^{0}(\cos \theta) \bigg],
\label{scattstate}
\end{equation}
where $f_0(\mathcal{E}_{\nu})$ is the $s$-wave scattering amplitude. 
To determine it, we first need to expand the scattering state in the angular region $\theta_0 < \theta < 1/\nu$, with $\nu \gg 1$, for which the associated Legendre functions can be approximated as \cite{NIST,zhang2018}: $P_{\nu}^{0}(\cos \theta) = 1 + \text{o}(\theta^2)$ and $Q_{\nu}^{0}(\cos \theta) = -\ln(\theta\nu e^{\gamma_{\text{E}}}/2) + \text{o}(\theta,\nu^{-1})$,
where $\gamma_{\text{E}}$ is the Euler-Mascheroni constant. 
Following the analogous procedure with respect to the flat 2D case \cite{landau,dalibard}, we define the length $a_s$, and thus the corresponding angle $\theta_s = a_s/R$, as the distance at which the scattered state $\Psi_{\nu}^{0}$ vanishes, i.~e.~$\Psi_{\nu}^{0}(\theta_s,\varphi) = 0$. 
After imposing this condition in Eq.~\eqref{scattstate}, in which the associated Legendre functions are expressed in the asymptotic form discussed above, we obtain 
\begin{equation}
f_0(\mathcal{E}_{\nu}) = -\frac{4}{\cot \delta_0(\mathcal{E}_{\nu}) -i},
\label{scatteringamplitude}
\end{equation}
which is the amplitude of $s$-wave scattering on the surface of a sphere, where the cotangent of the phase shift reads 
\begin{equation}
\cot \delta_0(\mathcal{E}_{\nu}) = \frac{2}{\pi}  \ln\bigg( \frac{\nu\, \theta_s \, e^{\gamma_{\text{E}}}}{2} \bigg)
\label{cotphaseshift}
\end{equation}
\cite{NIST,zhang2018}, which coincides with that of Ref.~\cite{zhang2018} if their constant $B$ is identified with $B^{-1} = \ln(R\, \theta_0/a_s)$. 
Let us then remark that our scattering amplitude is only proportional to the one calculated in Ref.~\cite{zhang2018}, and the different prefactor is chosen in such a way that $f_0(\mathcal{E}_{\nu})$ coincides with the effective interaction strength $\tilde{g}_{e,l_c}(\mathcal{E}_{l_0})$. 
This can be seen by comparing our scattering state in Eq.~\eqref{scattstate} with the analogous one for 2D scattering in Ref.~\cite{dalibard}, in which the same constants are chosen to ensure the validity of Eq.~\eqref{condition}.
Note that it is possible to choose the multiplicative prefactor in the scattering amplitude as required (see for instance the papers on 2D scattering \cite{lapidus1982,adhikari1986,landau,petrov2001}, and in particular the discussion in \cite{adhikari1986}), provided that the scattered wave is divided by the same factor, and the cross section is thus correctly calculated. 

As a last step, {we set $\nu=l_0$, and using the expressions obtained in Eqs.~\eqref{Tlc} and \eqref{scatteringamplitude} we impose the condition of Eq.~\eqref{condition}, i.~e.~that $\tilde{g}_{e,l_c} (\mathcal{E}_{l_0}) = f_0(\mathcal{E}_{l_0})$,} which allows us to identify $\tilde{g}_0$. Reintroducing the dimensional constants as $g_0 = \hbar^2 \tilde{g}_0/m$, we obtain the zero-range interaction strength:
\begin{equation}
g_0 = - \frac{2\pi\hbar^2}{m} \frac{1}{\ln\big[\sqrt{l_c(l_c+1)}\, a_s e^{\gamma_{\text{E}}}/(2R)\big]}, 
\label{g0rensphere}
\end{equation}
which is expressed as a function of the cutoff $l_c$, of the $s$-wave scattering length on the sphere $a_s$, and of the radius $R$. 
This is a central result of this paper, and it will be crucial for deriving the regularized equation of state.

\section{Equation of state of a spherical two-dimensional Bose gas}
\label{III}
We derive here the grand canonical potential $\Omega$ of a bosonic gas confined on the surface of a sphere. 
In particular, we calculate it as $\Omega = -\beta^{-1} \ln \mathcal{Z}$,
where $\beta=1/(k_{\text{B}}T)$ is the inverse temperature, with $k_{\text{B}}$ being the Boltzmann constant, and where $\mathcal{Z}$ is the grand canonical partition function. 
Following Ref.~\cite{tononi2019}, we calculate $\mathcal{Z}$ as a coherent-state functional integral over the bosonic field $\psi(\theta,\varphi,\tau)$ \cite{nagaosa}, namely, 
\begin{equation}
\mathcal{Z} = \int \mathcal{D}[\bar{\psi},\psi] \; 
e^{-\frac{S[\bar{\psi},\psi]}{\hbar}},
\end{equation}
where the Euclidean action is given by 
\begin{equation}
S[\bar{\psi},\psi] = \int_{0}^{\beta\hbar} \text{d}\tau \, 
\int_{0}^{2 \pi} \text{d} \varphi \, \int_{0}^{\pi} \text{d} \theta \, \sin\theta 
\, R^2 \,{ \cal L }(\bar{\psi},\psi), 
\label{actionsphere}
\end{equation}
which is written in terms of the Euclidean Lagrangian 
\begin{align}
{ \cal L } &= \bar{\psi}(\theta,\varphi,\tau) 
\bigg(\hbar\partial_{\tau}+\frac{\hat{L}^2}{2mR^2}-\mu + \Omega_z \hat{L}_z\bigg) 
\psi(\theta,\varphi,\tau) 
\nonumber
\\
&+ \frac{g_0}{2} |\psi(\theta,\varphi,\tau)|^{4},
\label{lagrsphere}
\end{align}
where $\mu$ is the chemical potential, $\tau$ is the imaginary time, and where the term $\Omega_z \hat{L}_z$, with $\hat{L}_z = -i \hbar\, \partial_{\varphi}$ being the angular momentum in the $z$ direction, models the additional energy contribution due to the rotation with angular velocity $\Omega_z$. 
{We stress that, in this formulation, the quartic term in the Lagrangian results from the integral
\begin{align}
\frac{g_0}{2} |\psi(\theta,\varphi,\tau)|^{4} = \frac{1}{2}\int_{0}^{2 \pi} \text{d} \varphi' \, \int_{0}^{\pi} \text{d} \theta' \, \sin\theta' \, R^2 \nonumber \\ \times   |\psi(\theta,\varphi,\tau)|^{2} E_R V_0(\theta',\varphi') |\psi(\theta+\theta',\varphi+\varphi',\tau)|^{2},
\end{align}
where the zero-range interaction in real space $V_0(\theta,\varphi)$ is defined after Eq.~\eqref{zerorangeint}, and here we multiply it by $E_R$ to reintroduce the correct dimension of energy.}

To calculate the grand potential, we implement the Bogoliubov-Popov theory by writing $\psi (\theta, \varphi, \tau) = \psi_0 + \eta (\theta, \varphi, \tau)$ and expanding the Lagrangian \eqref{lagrsphere} up to the Gaussian fluctuations of $\eta (\theta, \varphi, \tau)$. 
The Gaussian functional integral in $\eta$ can be calculated after expanding this field in the basis of $e^{i\omega_n \tau} \mathcal{Y}_l^{m_l}(\theta,\varphi)$ and then performing the sum over the Matsubara frequencies $\omega_n$. 
It yields \cite{tononi2019}
\begin{equation}
\Omega(\mu,\psi_0^2) = \Omega_{0}(\mu,\psi_0^2) +\Omega_{\text{g}}^{(0)}(\mu,\psi_0^2) + \Omega_{\text{g}}^{(T)}(\mu,\psi_0^2),
\label{effectivepotential}
\end{equation}
where $\Omega_0 = 4\pi R^2 \, (-\mu \psi_0^2 + g_0 \psi_0^4/2)$ is the mean-field grand potential. 
The Gaussian fluctuations with respect to the mean-field configuration, encoding the quasiparticle energies at zero temperature and at finite temperature, constitute the other contributions. 
In particular, we have the zero-temperature grand potential 
\begin{equation}
\Omega_{\text{g}}^{(0)}(\mu,\psi_0^2) = \frac{1}{2} \sum_{l=1}^{\infty} \sum_{m_l = -l}^l [E_l (\mu, \psi_0^2) -\epsilon_l - \mu],
\end{equation}
which includes the counterterms due to the convergence-factor regularization \cite{altland2010}. 
The finite-temperature grand-potential contribution reads 
\begin{equation}
\Omega_{\text{g}}^{(T)}(\mu,\psi_0^2) = \frac{1}{\beta} \sum_{l=1}^{\infty} \sum_{m_l = -l}^l \ln \{1-e^{-\beta[E_l (\mu, \psi_0^2) + m_l \hbar \Omega_z]}\},
\end{equation}
where we define the excitation spectrum as 
\begin{equation}
E_{l}(\mu,\psi_0^2)=\sqrt[]{(\epsilon_l
-\mu + 2 g_0 \psi_{0}^2 )^{2}-g_0^2 \psi_{0}^4},
\label{effectivespectrumsphere}
\end{equation}
with $\epsilon_l = \hbar^2 l(l+1)/(2m R^2)$. 
The effective grand potential in Eq.~\eqref{effectivepotential} depends on the parameter $\psi_0$, which is fixed by imposing the saddle-point condition $0=\partial \Omega/\partial \psi_0$, which allows us to identify the condensate density as $n_0(\mu) = \psi_0^2$. 
It reads \cite{note}
\begin{equation}
n_0(\mu) = \frac{\mu}{g_0} -
\frac{1}{4 \pi R^2} \sum_{l=1}^{\infty} 
\sum_{m_{l}=-l}^{l} \, \frac{2\epsilon_l + \mu}{E_{l}^{\text{B}}} \bigg(\frac{1}{2} + \frac{1}{e^{\beta E_{l}^{\text{B}}}-1} \bigg),
\label{n0}
\end{equation}
where the Bogoliubov spectrum is defined as
\begin{equation}
E_{l}^{\text{B}}=\sqrt[]{\epsilon_l 
(\epsilon_l +  2 \mu ) },
\label{Bogoliubovspectrumsphere}
\end{equation}
and it is obtained by considering the Gaussian terms as perturbative corrections with respect to $\psi_0^2 \approx \mu/g_0$. 
Substituting $\psi_0^2 = n_0(\mu)$ into the grand potential (see Refs.~\cite{tononi2019,note} for additional details), we get
\begin{align}
\Omega_0[\mu,n_0(\mu)] &= - (4 \pi R^2) \, \frac{\mu^2}{2 g_0},
\label{eq1}
\\
\Omega_{\text{g}}^{(0)}[\mu,n_0(\mu)] &=
\frac{1}{2} \sum_{l=1}^{\infty} 
\sum_{m_{l}=-l}^{l} \, (E_{l}^{\text{B}} -\epsilon_l - \mu),
\label{eq2}
\\ 
\Omega_{\text{g}}^{(T)}[\mu,n_0(\mu)] &= 
\frac{1}{\beta}  
\sum_{l=1}^{\infty} \sum_{m_{l}=-l}^{l}  \ln[1-e^{-\beta (E_{l}^{\text{B}}+ m_l \hbar \Omega_z)}],
\label{eq3}
\end{align}
which are the contributions of the grand potential of a spherical Bose gas under rotation.

\subsection{Grand potential and number density}
\label{IIIA}
Let us now obtain the regularized equation of state of the spherical Bose gas by using the previous results of scattering theory. 
To study the thermodynamics of the system, we need to set the angular rotation of the superfluid film to $\Omega_z = 0$, and the consequences of $\Omega_z \neq 0$ for the transport properties will be analyzed in the following section. 

The zero-point energy fluctuations of the Bose gas, encoded in the contribution of Eq.~\eqref{eq2}, are ultraviolet divergent as a consequence of the simplified zero-range interaction that we consider \cite{popov1972}. 
To regularize this term, we include the cutoff $l_c$ in the sum over $l$ and express the zero-temperature grand potential per unit of area as
\begin{equation}
\frac{\Omega^{(0)}}{4\pi R^2} = -\frac{\mu^2}{2 g_0}+
\frac{1}{8 \pi R^2} \sum_{l=1}^{l_c} 
(2l+1) \, (E_{l}^{\text{B}} -\epsilon_l - \mu),
\label{Omega0sumlc}
\end{equation} 
where $l_c {\gg 1}$ is unknown. 
To calculate the sum over $l$ of the function $\omega(l)=(2l+1) (E_{l}^{\text{B}} -\epsilon_l - \mu)$, we adopt the Euler-Maclaurin formula truncated at the second order:
\begin{equation}
\sum_{l=1}^{l_c} \omega(l) = \int_1^{l_c} \omega(l) \, \text{d}l - \frac{1}{2} \sum_{l=1}^{l_c} \frac{d \omega}{dl}  - \frac{1}{6} \sum_{l=1}^{l_c} \frac{d^2 \omega}{dl^2},
\end{equation}
which provides additional corrections with respect to the direct substitution of the sum with an integral, and which neglects additional higher-order terms in the formula, which are smaller than $\text{o}[(\xi/R)^2]$, with $\xi=\sqrt{\hbar^2/(2m\mu)}$ being the healing length. 
While the integral term can be calculated analytically, the summations must be evaluated numerically for $l_c \gg 1$, and since the latter depend on the ratio $\mu/E_R$, it is necessary to fix this parameter. 
In practice, we expect that the typical experiments \cite{tononi2021} are done in the regime of $\mu/E_R \in [1,10^3]$, in which these sums scale almost linearly with the chemical potential $\mu$, with a coefficient determined numerically. 

After we implement the procedure discussed here, Eq.~\eqref{Omega0sumlc} yields
\begin{align}
\frac{\Omega^{(0)}}{4 \pi R^2} =&- \frac{\mu^2}{2 g_0} -\frac{m\mu^2}{8\pi\hbar^2} \ln \bigg[ \frac{\hbar^2 l_c (l_c+1)}{m(E_1^B +\epsilon_1 +\mu) R^2 \, e^{1/2}}  \bigg] \nonumber
\\
&+\frac{m E_1^B}{8\pi\hbar^2} (E_1^B -\epsilon_1 - \mu) - {\frac{C_1 \mu}{4 \pi R^2}}, 
\end{align}
with $C_1=0.62$ being 
a numerical coefficient, and
which is clearly divergent in the limit of  $l_c \to \infty$. 
This logarithmic divergence, however, is perfectly counterbalanced by the analogous logarithmic one of $g_0$ [see Eq.~\eqref{g0rensphere}]. 
In conclusion, substituting $g_0$ in the zero-temperature grand potential, we get 
\begin{align}
\frac{\Omega^{(0)}}{4 \pi R^2}  = -\frac{m\mu^2}{8\pi\hbar^2} \bigg\{& \ln \bigg[ \frac{4\hbar^2}{m(E_1^B +\epsilon_1 + \mu) a_s^2 \, e^{2\gamma+1}}\bigg] +\frac{1}{2} 
\nonumber
\\
& {+ \frac{2 C_1\hbar^2}{m \mu R^2}} \bigg\}+\frac{m E_1^B}{8\pi\hbar^2} (E_1^B -\epsilon_1 - \mu),
\label{omegasphere}
\end{align}
which is the zero-temperature grand potential of a spherical superfluid film.
We stress that, in the limit of infinite radius $R\to \infty$, we have $\epsilon_1,E_1^{B} \to 0$, and our equation of state coincides with the standard one-loop result of Ref.~\cite{mora2003}. At the same time, our result includes the leading curvature corrections to the equation of state of a large spherical Bose gas. 

\begin{figure}
\centering
\includegraphics[width=\columnwidth]{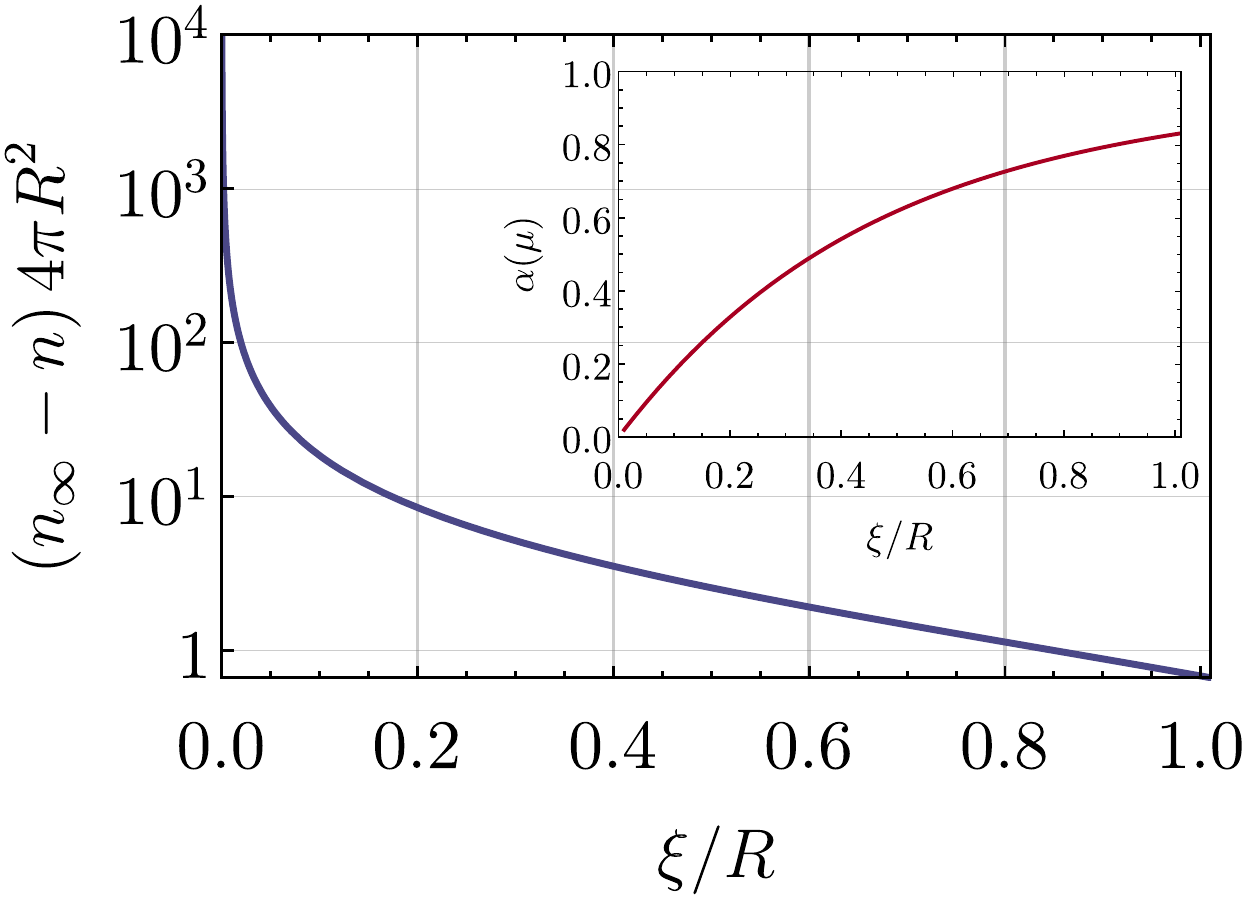}
\caption{Finite-size corrections to the zero-temperature equation of state of a spherical Bose gas of density $n$, given by Eq.~\eqref{Nsphere} at $T=0$. We show in particular the adimensional quantity $(n_{\infty}-n) \, 4 \pi R^2$, where $n_{\infty}$ is the density of the infinite system \cite{mora2003}, obtained from Eq.~\eqref{Nsphere} in the limit of $R \to \infty$, as a function of the parameter $\xi/R$. The inset shows the behavior of the function $\alpha(\mu)$ [see Eq.~\eqref{alfa}], which tends to zero at the thermodynamic limit. }
\label{fig1}
\end{figure}

The number density $n=N/(4\pi R^2)$ of the spherical Bose gas can be calculated by deriving the grand-potential density with respect to the chemical potential, i.~e.~$n = - \partial[\Omega/(4\pi R^2)]/\partial \mu$. Including also the finite-temperature contribution, which originates from $\Omega_{\text{g}}^{(T)}$, we get 
\begin{align}
n &= 
\frac{m \mu}{4\pi\hbar^2} \, \ln \bigg\{ \frac{4\hbar^2 [1-\alpha(\mu)]}{m\mu\, a_s^2 \, e^{2\gamma+1+\alpha(\mu)}}\bigg\} + {\frac{C_1}{4 \pi R^2}}
\nonumber \\
&+  \frac{1}{4\pi R^2} \, \sum_{l=1}^{\infty} \sum_{m_l=-l}^{l} \, \frac{\epsilon_l}{E_l^B} \frac{1}{e^{\beta E_l^B}-1},
\label{Nsphere} 
\end{align}
where we introduce the function 
\begin{equation}
\alpha(\mu) = 1-\frac{\mu}{\mu+E_1^B +\epsilon_1},
\label{alfa}
\end{equation}
which is positive and nonzero for finite values of the radius, and which tends to zero in the thermodynamic limit (see the inset in Fig.~\ref{fig1}). 
Our equation of state \eqref{Nsphere} represents the density of the spherical Bose gas at finite temperature, obtained at a Gaussian beyond-mean-field level, and it includes the corrections (with respect to the flat-case result \cite{mora2003}) due to the finite radius of the sphere. 
This result provides a refined description of the thermodynamics of a shell-shaped condensate and elucidates the quantitative role of curvature, controlled by the finite radius, in the spherically symmetric case. 

We illustrate the quantitative relevance of the finite-size corrections in Fig.~\ref{fig1}, where we plot the difference between the zero-temperature density of the infinite system $n_{\infty}$ and that of the finite sphere, multiplied by the area of the sphere itself $4\pi R^2$.
In particular, our theory is valid in the region of small $\xi/R$, in which these corrections to the number of particles in the system are more relevant. 
To observe these effects, it is thus favorable to have a sufficiently small $\xi/R$ (i.e., a large radius of the sphere), keeping in mind that the correction to the number density, which is the only well-defined quantity in the thermodynamic limit, vanishes in the infinite-radius regime of $\xi/R\to 0$. 
As a practical example, we consider a spherical gas with a sufficiently low density and assume that it is possible to tune the interatomic interactions, so that the ratio $\xi/R$ can be diminished while keeping the radius small and fixed. Then, for the realistic value of $\xi/R =1/50$ (see Ref.~\cite{tononi2021}), the expected difference between the number density and the infinite system one is quite relevant, since we find $(n_{\infty}-n) \, 4 \pi R^2 \approx 10^2$. 

\subsection{Microscopic derivation of the superfluid density}
\label{IIIB}
When a quantum liquid is rotated, for instance imposing the motion of the external potential, only the normal part of the fluid is dragged. 
In the Lagrangian in Eq.~\eqref{lagrsphere}, the additional energy contribution due to the trap rotation is encoded into the term $\propto \Omega_z \hat{L}_z$, where the angular velocity $\Omega_z$ is essentially a Lagrange multiplier. 
The average angular momentum density of the normal fluid $L_n$ can then be calculated by simply deriving the effective potential of Eq.~\eqref{effectivepotential} with respect to $\Omega_z$. 
Specifically, we set \cite{svistunov2015,tononi2018}
\begin{equation}
L_n = - \bigg( \frac{\partial \Omega (\mu, \psi_0^2)}{\partial \Omega_z} \bigg) \bigg\rvert_{\psi_0^2 = n_0(\mu)}, 
\end{equation}
and inspired by the approach of Landau \cite{landau1941}, we expand the angular momentum for a small rotational velocity $\Omega_z$. We find that
\begin{equation}
L_n = 
\beta \sum_{l=1}^{\infty} \sum_{m_l = -l}^l \hbar^2 m_l^2 \, \frac{e^{\beta E_l^{\text{B}} }}{(e^{\beta E_l^{\text{B}} }-1)^2} \,  \Omega_z,
\end{equation}
where the Bogoliubov spectrum $E_l^{\text{B}}$ appears when the condition $\psi_0^2 = n_0(\mu)$ is imposed. 
Considering that \cite{grads}
\begin{equation}
\sum_{m_l = -l}^l m_l^2 = \frac{1}{3} (2l+1) (l^2+l), 
\end{equation}
the angular momentum can be expressed as 
\begin{equation}
L_n = I_n \Omega_z, 
\end{equation}
where 
\begin{equation}
I_n = 
\frac{\beta}{3} \sum_{l=1}^{\infty} (2l+1) \, \hbar^2 (l^2+l) \, \frac{e^{\beta E_l^{\text{B}} }}{(e^{\beta E_l^{\text{B}} }-1)^2}  
\end{equation}
is the moment of inertia of the normal fluid. 
A spherical surface with total mass $M_n = m n_n^{(0)} (4 \pi R^2)$, where $n_n^{(0)}$ is the density of the normal fluid has an angular momentum given by $I_n = 2 M_n R^2/3$, which allows us to immediately identify the normal density $n_n^{(0)}$. 
The superfluid density is simply given by the difference between the total density and the normal density, namely
\begin{equation}
n_s^{(0)} = n - \beta \sum_{l=1}^{\infty} \frac{(2l+1)}{4\pi R^2} \, \frac{\hbar^2 l (l+1)}{2 m R^2} \, \frac{e^{\beta E_l^{\text{B}} }}{(e^{\beta E_l^{\text{B}} }-1)^2},
\label{ns0}
\end{equation}
which is exactly the result postulated in our previous work \cite{tononi2019}. 
However, we emphasize that Eq.~\eqref{ns0} is obtained here with a microscopic derivation based on the functional integral formulation of quantum field theory. 

We also stress that, in the limit of infinite area, our bare superfluid density coincides with that of a two-dimensional flat superfluid. Let us show this explicitly: we first substitute the sum in Eq.~\eqref{ns0} with an integral, which is exact when the spacing between the energy levels, proportional to $R^{-2}$, becomes zero. Then, identifying $4\pi R^2$ with the flat system area $L^2$, and redefining the integration variable with $k = [4\pi (l^2+l)/L^2]^{1/2}$, we get
\begin{equation}
n_{s, \, \text{flat}}^{(0)} = n - \beta \int_{\sqrt{8\pi}/L}^{\infty} \frac{\text{d}k}{2\pi} \, k \, \frac{\hbar^2 k^2}{2 m} \, \frac{e^{\beta E_k^{\text{B}} }}{(e^{\beta E_k^{\text{B}} }-1)^2},
\label{nsflat}
\end{equation}
with $E_k^{B}=\sqrt{\hbar^2k^2/(2m)[\hbar^2k^2/(2m)+2\mu]}$ being the flat-case Bogoliubov spectrum. 
It can be verified that, in the thermodynamic limit, Eq.~\eqref{nsflat} coincides with the superfluid density of a flat superfluid \cite{landaustat2}. 

\vspace{1mm}
\section{Conclusion} 
We analyzed the low-energy scattering problem on a spherical surface, and we expressed the interaction strength of a zero-range potential in terms of the $s$-wave scattering length. 
With this crucial relation, we obtained the regularized equation of state of a spherical Bose gas at finite temperature, and we provide a microscopic derivation of the superfluid density of the system. 
Our results will support experiments with two-dimensional shell-shaped Bose-Einstein condensates obtained by trapping a quantum gas with radio-frequency-induced adiabatic potentials. 
Even if the cleanest environment for implementing these traps is constituted by space-based microgravity facilities such as the Cold Atom Lab \cite{carollo2021,elliott2018,aveline2020,frye2021},
the experiments can also be conducted in earth-based microgravity laboratories \cite{condon2019}. 
In principle, by limiting the investigation to sufficiently small shells and by properly tuning the quadrupolar magnetic field, it is possible to test our predictions also in standard ground-based laboratories \cite{guo2021}.
An alternative promising path for the realization of shell-shaped condensates is offered by bichromatic optical dipole traps \cite{onofrio2004,mas2019,wilson2021}, combined with a magnetic field gradient to counterbalance gravity. 
The variety of possible implementations will allow us to investigate our finite-temperature equation of state and 
to explore the emergent physics of vortices, long-range interactions, and other nontrivial phenomena \cite{shi2017,moller2020} in a curved quantum gas with the topology of a sphere.  

\begin{acknowledgments}
I acknowledge useful discussions with A. Pelster, D. Petrov, and L. Salasnich, and thanks T.-L. Ho for useful remarks. R. Onofrio is acknowledged for pointing out the possible implementation of bichromatic optical dipole traps. 
This research has been supported by the ANR Grant Droplets (19-CE30-0003).
\end{acknowledgments}



\begin{thebibliography}{50}
\bibitem{note2} One may consider the summation of the Born series as a ``renormalization" of the bare interaction strength $g_0$: the renormalized interaction is then given by $\tilde{g}_e (\mathcal{E}_{l_0})$, and this procedure will allow us to express $g_0$ in terms of the scattering length and of the cutoff $l_c$.
\end{thebibliography}
\end{document}